\documentclass[superscriptaddress,twocolumn]{revtex4-1}

\usepackage[utf8]{inputenc}
\usepackage{amsmath}
\usepackage{amssymb}
\usepackage{natbib}
\usepackage{graphicx}
\usepackage{siunitx}
\usepackage{xcolor}
\usepackage{textcomp}

\usepackage{xr}
\begin{document}
\title{Thresholdless transition to coherent emission at telecom wavelengths from coaxial nanolasers}

\author{S\"oren Kreinberg}
\affiliation{Technische Universit\"at Berlin, Institut f\"ur Festk\"orperphysik, Hardenbergstr.~36, 10623 Berlin, Germany}	
\author{Kaisa Laiho}
\affiliation{Technische Universit\"at Berlin, Institut f\"ur Festk\"orperphysik, Hardenbergstr.~36, 10623 Berlin, Germany}
\author{Frederik Lohof}
\affiliation{Universität Bremen, Institut f\"ur Theoretische Physik, Otto-Hahn-Allee 1, 28359 Bremen, Germany}
\author{William E. Hayenga}
\affiliation{CREOL, The College of Optics and Photonics, University of Central Florida, Orlando, FL, USA}
\author{Pawe\l{}~Holewa}
\affiliation{Technische Universit\"at Berlin, Institut f\"ur Festk\"orperphysik, Hardenbergstr.~36, 10623 Berlin, Germany}
\author{Christopher Gies} 
\email{Corresponding author: gies@itp.uni-bremen.de}
\affiliation{Universität Bremen, Institut f\"ur Theoretische Physik, Otto-Hahn-Allee 1, 28359 Bremen, Germany}
\author{Mercedeh Khajavikhan}
\email{Corresponding author: mercedeh@creol.ucf.edu}
\affiliation{CREOL, The College of Optics and Photonics, University of Central Florida, Orlando, FL, USA}
\author{Stephan Reitzenstein}
\email{Corresponding author: stephan.reitzenstein@physik.tu-berlin.de}
\affiliation{Technische Universit\"at Berlin, Institut f\"ur Festk\"orperphysik, Hardenbergstr.~36, 10623 Berlin, Germany}
\begin{abstract}
\end{abstract}

\maketitle
\textbf{The ongoing miniaturization of semiconductor lasers has enabled ultra-low threshold devices~\cite{Oulton2007,Ellis2011} and even provided a path to approach thresholdless lasing with linear input-output characteristics~\cite{Khajavikhan2012,Prieto2015,Ota2017}. Such nanoscale lasers have initiated a discourse on the origin of the physical mechanisms involved and their boundaries, such as the required photon number~\cite{Mayer2016, Mork2018}, the importance of optimized light confinement in a resonator~\cite{Lermer2013} and mode-density enhancement~\cite{Strauf2006, Hayenga2016}. Here, we investigate high-$\beta$ metal-clad coaxial nanolasers, which facilitate thresholdless lasing. We experimentally and theoretically investigate both the conventional lasing characteristics, as well as the photon statistics of the emitted light. While the former lacks adequate information to determine the threshold to coherent radiation, the latter reveals a finite threshold pump power. Our work clearly highlights an important and often misunderstood aspect of high-$\beta$ lasers, namely that a thresholdless laser does have a finite threshold pump power and must not be confused with a hypothetical  zero threshold laser.}

Thresholdless input-output curves have been observed in metallic coaxial nanolasers (CNLs) with InGaAsP quantum wells (QWs) as active material \cite{Khajavikhan2012, Hayenga2016a}. These structures emit radiation at telecommunication wavelengths and offer high potential for applications in silicon photonics and integrated optics \cite{Chen2011, Zhu2018}. In particular, they can be fabricated with outer diameters of only a few hundred nanometers, resulting in footprints that are considerably smaller than for their dielectric thresholdless laser counterparts---so far reported only in quantum-dot systems \cite{Prieto2015,Takiguchi2016,Ota2017}. The possible observation of thresholdless lasing in QW systems brings about a debate as to whether these devices are indeed lasers capable of generating coherent radiation. In addition, it opens up a course of discussions on the role of various physical mechanisms required to ensure operation in this regime.

While the properties and the terminology of thresholdless lasing have long accompanied the development of laser physics, the much more recent realization of nanoscale devices operating in the thresholdless regime have brought these issues again to the forefront. Nanolasers manifesting close-to-linear input-output characteristics in a double-logarithmic scale are usually referred to as \emph{thresholdless lasers} \cite{Bjoerk1994}, in which case all the spontaneously generated photons are emitted into the lasing mode ($\beta = 1$). The disappearance of the intensity jump is a consequence of the absence of radiative and non-radiative emission losses in high-$\beta$ lasers. However, the term thresholdless, which refers to a  linear input-output characteristics, is somewhat ambiguous, as such lasers may wrongly be understood to exhibit a zero threshold pump power. To resolve this issue, it is paramount to identify the excitation level at which the onset of the stimulated emission occurs via quantum optical studies. The autocorrelation function is the measurement of choice, as it reveals a distinctive change from chaotic to Poissonian photon statistics when coherent emission is reached. Alternatively, true photon-number resolving detectors can be used to directly examine the photon statistics \cite{Dynes2011, Schlottmann2017}. When dealing with high-$\beta$ nanolasers operating in the cavity quantum-electrodynamically enhanced regime, one needs to carefully assess both the emission intensity and the statistical properties of the emitted light, as has been pointed out in the demonstration of nanolasers operating with different gain materials and cavity designs ~\cite{Strauf2006, Wiersig2009, Jagsch2018,Chow2018}. This is of particular relevance for a new generation of nanolasers using monolayer flakes of semiconducting transition-metal dichalcogenides (TMDs) as gain material in high-Q resonators with localized modes, see e.g.~\cite{Wu2015, Ye2015,Javerzac-Galy2018}. In the vivid development of utilizing TMDs for optoelectronics, the observation of ultra-low threshold lasing has been claimed in several publications solely based on the spectral emission properties. However, the importance of measuring the autocorrelation function has been pointed out e.g. in Ref.~\cite{Reeves2018}. Importantly, our studies of nanolasers with close-to thresholdless behavior do show the existence of a finite threshold, clearly highlighting that thresholdless lasing should not be confused as lasing with a zero threshold pump power. In fact, the latter could only be achieved in a hypothetical cavity subject to no losses.

In this work, we measure for the first time the power dependent second-order coherence for two metallic nanolasers operating close to the thresholdless lasing regime. We compare conventional signs of lasing, observed in a standard micro-photoluminescence (\textmu{}-PL) setup, to quantum-optical studies of the second-order autocorrelation function. Accompanying the experimental investigation, we employ a microscopic laser model to obtain important insight in the interplay of the CNL's properties and their effect on the photon statistics of the emission. The theory is based on a quantum-optical treatment of the charge carriers and their interaction with the quantized light field, providing simultaneous access to the input-output characteristics, the carrier population functions, the first- and second-order photon correlations and, thereby, to the power-dependence of the coherence time and the second-order autocorrelation function $\smash{g^{(2)}(\tau)}$ (for more details see the Supplementary Information (SI)). Importantly, in contrast to conventional microlaser theories, the $\beta$-factor does not play the role of a constant device-dependent parameter, but is calculated directly from the spontaneous-emission (SE) rates of carriers in the band structure into lasing and non-lasing modes. We find the resulting $\beta$-factor to be excitation-power dependent due to phase-space filling effects. Additionally, we show that special care has to be taken when investigating the second-order correlation function of nanolasers by using narrow-band spectral filtering. Indeed, in our case it leads into an enhanced intensity noise resulting in pseudo-thermal characteristics, which  can cause pitfalls in the interpretation of the emitted light \cite{Neelen1992, Neelen1993}.
 


\begin{figure}[t]
\centering
\includegraphics[width=1.0\columnwidth]{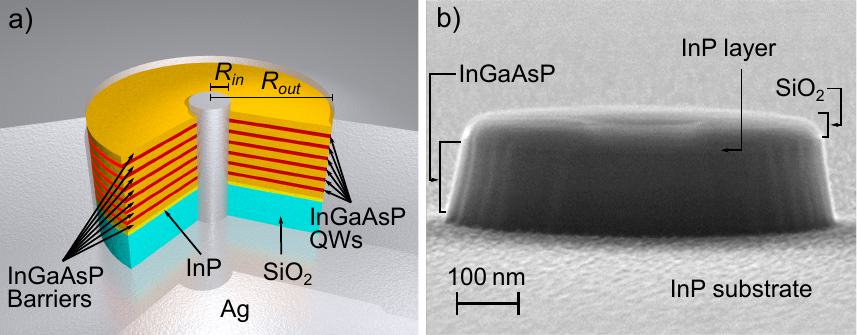}
\caption{(a) Cut-away schematic of metallic coaxial nanolaser CNL\,1. (b) Scanning electron microscope image (SEM) of the dielectric filling of a CNL after dry etching, before silver deposition. After silver deposition the sample is flipped and the InP substrate is etched away. A \SI{10}{\nano\meter} capping layer of InP covers the InGaAsP region. The ends of the active medium are terminated by a \SI{100}{\nano\meter}-high $\mathrm{SiO}_2$ plug on one side and a \SI{20}{\nano\meter}-high vacuum plug on the other, which provides an avenue to pump the devices and to also outcouple the generated radiation. Details on the fabrication can be found in the SI.}
\label{fig:sample}
\end{figure}

The investigated coaxial nanolasers are depicted in Fig.~\ref{fig:sample}. The active medium in these cavities comprises a ring of six InGaAsP QWs with an overall height of \SI{200}{\nano\meter} radially sandwiched between an inner silver core with a radius of $R_\mathrm{in}$ and an outer metallic cladding confining the active medium to a radius of $R_\mathrm{out}$. The two CNLs under study share a similar geometry, differing only in their inner and outer radii. This mainly alters the spectral locations of the supported modes and, in return, also the effective $\beta$-factor.  The QWs result in a broad gain spectrum that spreads over several tens of nanometers close to the telecom E- and S-bands at cryogenic temperatures.
We label the investigated structures with CNL1 ($R_\mathrm{out}$: \SI{295}{\nano\meter}, $R_\mathrm{in}$: \SI{55}{\nano\meter}) and CNL2 ($R_\mathrm{out}$: \SI{315}{\nano\meter}, $R_\mathrm{in}$: \SI{75}{\nano\meter}). In both cases the azimuthally polarized $\mathrm{TE}_{01}$-like mode, exhibiting  the largest quality factor and highest confinement factor, is expected to lase. A simulation of the optical modes and their properties is presented in the SI.


\begin{figure}[t]
\centering
\includegraphics[width=1.0\linewidth]{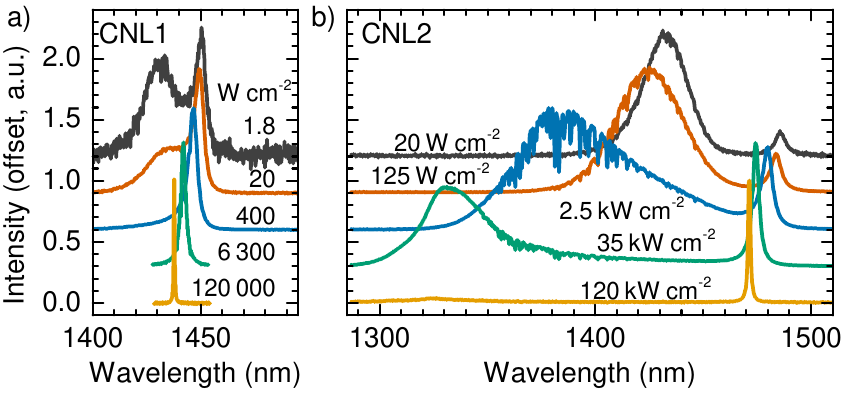}
\caption{Excitation intensity-dependent $\mu$-PL spectra for both CNLs. At low excitation intensities a broad non-lasing spectral peak is found at smaller detuning for CNL1 ($\delta \lambda$ = 20\,\SI{}{\nano\meter}) than for CNL2  ($\delta \lambda$ = 53\,\SI{}{\nano\meter}). In all spectra, the height of the most prominent peak is normalized to unity. Due to absorption in humid air distortion is apparent in some spectra. The broad emission feature observed for both CNLs in the shorter wavelengths is associated with the $\mathrm{TM}_{01}$-like mode. The gain maximum is centered at about 1430 nm and shows only a slight dependence of the excitation intensity (see SI; Fig.~S7).}
\label{fig:gainspectra}
\end{figure}
\begin{figure}[t]
\centering
\includegraphics[width=0.95\linewidth]{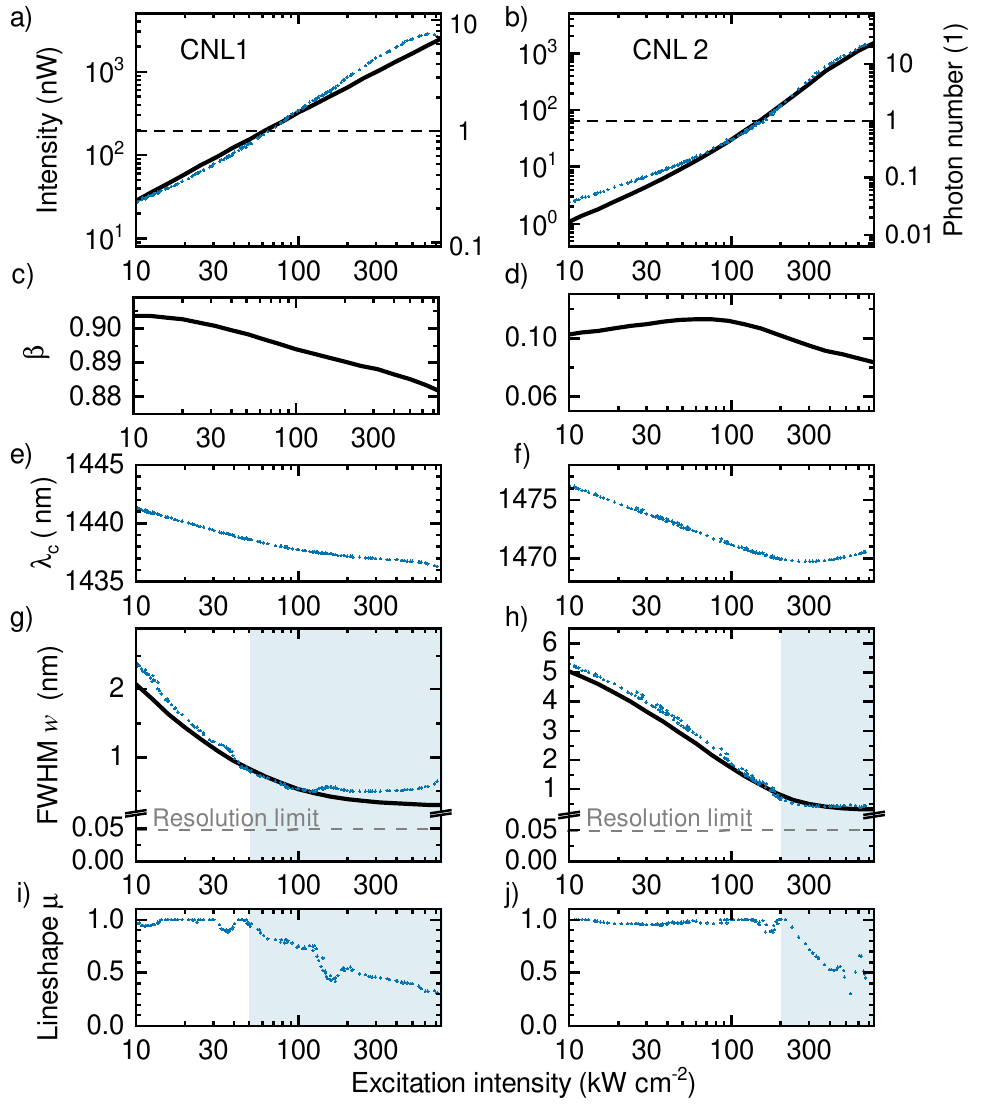}
\caption{Excitation power dependent optical properties of the CNLs: (a-b) integrated emission intensity and intra-cavity photon number, (c-d) $\beta$-factor, (e-f) central emission wavelength $\lambda_{c}$, (g-h) spectral full width at half maximum (FWHM) $w$ and (i-j) lineshape parameter $\mu$ for CNL1 (left) 
 and CNL2 
(right).  Measured values are illustrated with symbols, whereas the black solid lines are predicted by the theoretical model (see SI). The dashed lines in (a-b) indicate a mean intra-cavity photon number of one, which is often regarded as a threshold condition for nanolasers \cite{Bjoerk1994, Reeves2018}. CNL1 and CNL2 meet this criterion at pump rates of approximately \SI{65}{\kilo\watt\per\square\centi\meter} and \SI{155}{\kilo\watt\per\square\centi\meter}, respectively.
We note that the azimuthally polarized light from the CNLs is projected to a linear polarization prior to measurement. The wiggling artifacts in (e-f) are caused by absorption of the CNL emission in humid air. }
\label{fig:input-output}
\end{figure}

Preliminary information about laser action in the CNLs is obtained by excitation-intensity dependent \textmu{}-PL-studies. The experimental setup used for our investigations is depicted in the SI. A selection of the measured spectra is presented in Fig.~\ref{fig:gainspectra} for both CNLs.
To extract the conventional laser characteristics, which are shown in Fig.~\ref{fig:input-output}, the measured \textmu{}-PL spectra are fitted with pseudo-Voigt profiles (see Methods). While the integrated mode intensity of CNL1 (Fig.~\ref{fig:input-output}a) features a nearly linear input-output curve resembling thresholdless lasing, that of CNL2 (Fig.~\ref{fig:input-output}b) 
shows a more prominent S-shape indicating a reduced $\beta$-factor.
The experimental input-output curves agree well with our quantum-optical laser model (solid black curves in Fig.~\ref{fig:input-output}) apart from the slight deviation from the S-shape of CLN2, which is explained by the possible contribution of zero-dimensional gain centers at low excitation ~\cite{Jagsch2018} and is not captured by a description based solely on a two-dimensional QW gain medium.

Spontaneous and stimulated emission into the laser mode result from the steady-state occupation of carrier states in the band structure. For our laser model, the light-matter coupling constants for both CNLs are determined by matching the measured and calculated cavity-enhanced time-resolved photoluminescence (TRPL) traces (see SI; Fig.~S4). Radiative losses enter in the form of an effective rate, which then allows us to calculate the $\beta$-factor (see SI;  Eq.~(S24)). As shown in Fig.~\ref{fig:input-output}c,d it exhibits a decrease with increasing excitation power. Several factors have been identified to modify the SE rate and $\beta$-factor at constant excitation power, such as line-shape effects and spectral detuning~\cite{Fujita2001,Englund2005}, even in the presence of fast dephasing~\cite{Sumikura2016}, as well as population effects \cite{gregersen_quantum-dot_2012}. 
Here, the origin of the observed excitation power-induced change of the $\beta$-factor is attributed to the non-equilibrium carrier distribution functions that enter the SE rate and, thereby influence $\beta$.  At the onset of stimulated emission, we observe hole burning in the carrier distribution functions in spectral vicinity of the cavity resonance. With increasing excitation power, populations still rise in parts of the band structure that are not depleted by stimulated emission into the laser mode. The SE rate in the laser mode no longer grows, as the populations around the mode are fixed by the hole-burning effect. The increasing number of carriers in higher-lying states leads to a larger SE into the non-lasing modes, which in effect lowers the $\beta$-factor (see SI). The peak value of $\beta$ is reached at the point of the maximum SE rate into the laser mode in relation to radiative losses. The higher values of $\beta$ with a maximum of 0.9 obtained for CNL1 are consistent with its practically thresholdless behavior. In CNLs the high $\beta$-factor stems from a strong light matter interaction leading to an enhancement of radiative decay \cite{Hayenga2016a}. To support this explanation, we discuss the excitation-power dependence of $\beta$ in the SI in more detail. In this context, we present TRPL lifetime measurements and calculations for both CNLs and compare these with the lifetime in planar material, from which a Purcell factor of about 15 is estimated.


A closer inspection of the spectral characteristics of the CNLs is required to access other conventional signs of lasing. Fig.~\ref{fig:input-output}e,f reveal that the central wavelengths of both CNLs show strong blue-shifting with respect to the growing excitation intensity. However, at higher intensities, CNL2's 
emission  wavelength experiences red-shifting, which is attributed to heating effects. Additionally, both CNLs show linewidth narrowing by about one order of magnitude for the excited mode within the measured intensity range, which is in agreement with the calculations presented in Fig.~\ref{fig:input-output}g,h. Within the investigated excitation-intensity range the spectral lineshapes change from a Lorentzian to Gaussian profile, which is reflected in the Pseudo-Voigt profile by $\mu$ changing from 1 to about 0.3 (see Methods) in Fig.~\ref{fig:input-output}i,j. Interestingly, this change begins at the same excitation intensities at which the emission linewidth starts to saturate, that is around \SI{50}{\kilo\watt\per\square\centi\meter} for CNL1 and around \SI{200}{\kilo\watt\per\square\centi\meter} for CNL2, possibly indicating the onset of temperature induced inhomogeneous broadening at higher excitation intensities. While such thermal effects are not included in the modeling, the key saturation feature of the linewidth reduction is reproduced irrespective of the spectral lineshape. From the linewidths, coherence times can be obtained, and we estimate a lower limit of \SI{0.3}{\pico\second} below threshold to \SI{8}{\pico\second} above threshold (see Methods).

\begin{figure}[t]
\centering
\includegraphics{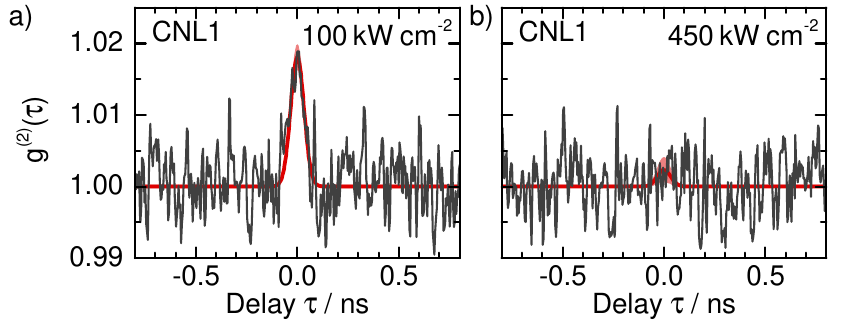}
\includegraphics{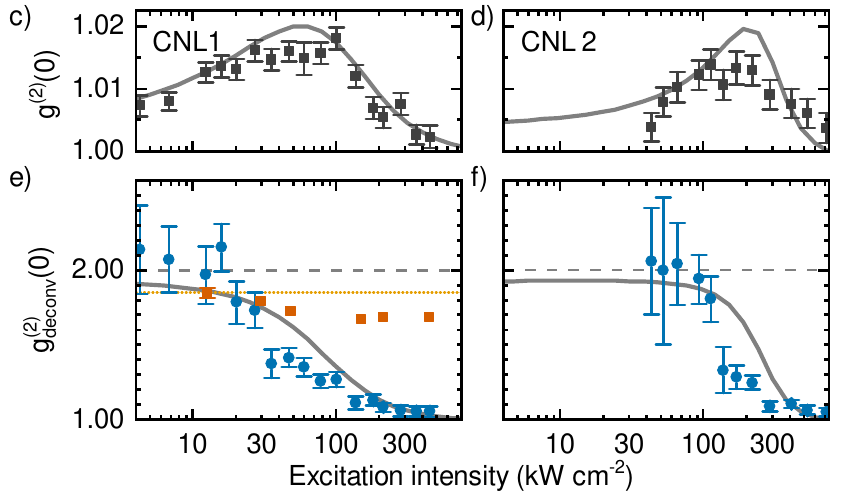}
\caption{(a-b) Experimental $g^{(2)}(\tau)$ coincidence histograms (black lines) for CNL1 
measured at different excitation intensities, together with the Gaussian fits (red lines, uncertainties pink) yielding $g^{(2)}(0)$ = 1.0180$\pm$0.0018 and 1.0023$\pm$0.0018, respectively. The temporal resolution of our HBT setup is about \SI{80}{\pico\second} corresponding to the FWHM of the measured correlogram in (a). (c-d) Raw (black squares) and (e-f) deconvoluted $\smash{g^{(2)}(0)}$ values (blue circles) for CNL1 (left) 
and CNL2 (right) 
with respect to excitation intensity. The theoretical predictions are illustrated with gray solid lines. The raw values in (c-d) fit well with the theoretical predictions in (e-f) after being convoluted with the estimated $\smash{g^{(2)}(\tau)}$-setup response time (See SI).  The red squares in (e) are raw values measured with narrow-band spectral filtering.  As a validation of our setup, the orange dashed line shows the value of $\smash{g^{(2)}(0)}$ measured for a spectrally filtered incandescent lamp, for which thermal characteristics is expected by definition. The gray dashed line provides a guide for the eye.}
\label{fig:g2}
\end{figure}

Recently, we showed that an almost linear input-output characteristic accompanied by linewidth narrowing can occur in the regime of amplified SE, which can be identified by assessing the photon statistics of the associated emission
~\cite{Kreinberg2017}. To rule out this possibility in the case of CNL1, we study the time-dependent second-order autocorrelation function $\smash{g^{(2)}(\tau)}$ (see Methods for details on the measurement) by selecting the CNLs' mode via a \SI{12}{\nano\meter} bandpass filter.
In Fig.~\ref{fig:g2}a,b we present two of the measured coincidence histograms $\smash{g^{(2)}(\tau)}$  for CNL1. 
We extract raw values of $\smash{g^{(2)}(0)}$ by fitting Gaussian temporal profiles to the measured data. The fitted peak heights are presented in Fig.~\ref{fig:g2}c,d for both CNLs showing gradually decreasing  $\smash{g^{(2)}(0)}$ at excitation intensities higher than about \SI{100}{\kilo\watt\per\square\centi\meter}. Below that, the raw values of  $\smash{g^{(2)}(0)}$ are more strongly limited by the temporal resolution of our  $\smash{g^{(2)}(\tau)}$-setup due to the excitation-power dependent coherence time \cite{Strauf2006, Ulrich2007}. The raw measured $\smash{g^{(2)}(\tau)}$ histograms are then deconvoluted, compensating for the finite timing-jitter (50 ps) of our superconducting nanowire single-photon detectors following the procedure introduced in Ref.~\cite{Kreinberg2017} (see Methods). In Fig.~\ref{fig:g2}e,f we present the deconvoluted $\smash{g^{(2)}_\mathrm{deconv}(0)}$ values for the investigated CNLs, which agree well with our  theoretical prediction.  Moreover, we perform a second series of excitation-intensity dependent measurements of $\smash{g^{(2)}(\tau)}$ for CNL1 with tight spectral filtering as shown in Fig.~\ref{fig:g2}e with red squares (see Methods). The effect of narrowband spectral filtering in the HBT measurement is of particular relevance when studying emission of nanolasers, which often have small cavity quality factors and large emission linewidths. In our case, a too narrow spectral selection in $\smash{g^{(2)}(\tau)}$-setup leads to artifacts in the associated photon auto-correlation and an incorrect interpretation of the measured autocorrelation function.

Altogether, the strong agreement between experiment and theory supports the validity of the deconvolution procedure performed on the raw $\smash{g^{(2)}(0)}$ data. Moreover, the raw data is reproduced by theory if the calculated coherence times and the time resolution of the $\smash{g^{(2)}(\tau)}$-setup are used to convolute the calculated result, producing the solid lines in  Fig.~\ref{fig:g2}c,d. We emphasize that these predicted lines are not obtained from a fit, but result from calculated coherence times and $\smash{g^{(2)}(0)}$ for a single consistent parameter set for each CNL (see SI).  Clearly, both the experiment and theory provide strong evidence for a transition from spontaneous to stimulated emission and indicate laser operation in both CNLs. The autocorrelation studies reveal that both CNLs have a wide transition region from spontaneous to stimulated emission spanning over two orders of magnitude of the output intensity \cite{Kreinberg2017, Lohof2018}.

In conclusion, our combined experimental and theoretical quantum-optical investigations of metal-clad coaxial nanolasers provide unambiguous evidence for thresholdless laser operation with coherent emission at telecom wavelengths. The validity of the first- and second-order coherence properties obtained from experiment is supported by  quantum-optical modeling, which yields simultaneous agreement of the input-output characteristics, coherence time, and autocorrelation function. From the microscopic laser model, we obtain a $\beta$-factor that is pump-power dependent in contrast to the wide-spread assumption of a device-dependent constant $\beta$.
Our results highlight that it is crucial to collect the spectrally broad emission line of a nanolaser without truncating its spectral tails in order to correctly access the underlying photon statistics. As such, this work provides the first comprehensive characterization of thresholdless lasing in high-$\beta$ metallic quantum-well nanoscale lasers. It paves the way for designing the next generation of ultrasmall coherent light sources, benefiting from quantum effects appearing at the nanoscale.


\section*{Methods}

\subsection*{Fitting the measured lineshapes }

For fitting the measured lineshapes we use pseudo-Voigt spectral profiles $I$ in terms of wavelength $\lambda$
given by \newline
$
I(\lambda) 
=A\left[\frac{\mu}{\pi}\frac{2w}{w^{2}+4(\lambda-\lambda_{\mathrm{C}})^{2}}+(1-\mu)\frac{2\sqrt{\ln2}}{w\sqrt{\pi}}e^{-4\ln2\frac{(\lambda-\lambda_{\mathrm{C}})^{2}}{w^{2}}}\right]
$
\newline
having the area $A$, the central wavelength $\smash{\lambda_{\mathrm{C}}}$, the spectral full-width of half maximum (FWHM) $w$ and the Lorentzian-Gaussian mixing parameter $\smash{\mu}$ ($0\le\smash{\mu}\le1$). For Lorentzian-shaped spectral lines, the coherence time is calculated from the frequency spectral FWHM $\Delta\nu$ by $\tau_\mathrm{coh, Lorentz}=(\pi\Delta\nu)^{-1}$ and can be approximated for a pseudo-Voigt lineshape by $\tau_{\mathrm{coh}}\approx(2-\mu)(\pi\Delta\nu)^{-1}$.

\subsection*{Measurement of the second-order autocorrelation function and deconvolution of experimental data}

The second-order autocorrelation function $g^{(2)}(\tau)$ can be accessed via the famous Hanbury Brown and Twiss (HBT) experiment by coincidence discrimination of emitted photons at the output ports of a symmetric beam splitter \cite{Brown1956}. Since HBT measurements can be conducted with single-photon sensitive click-detectors without true photon-number resolution, it has become the standard tool for classifying light sources via the time-resolved investigation of photon-correlations. As of now, measuring the second-order autocorrelation of light sources that have short coherence times sets great demands on the temporal resolution of the photo detection \cite{Tapster1998,Strauf2006,Ulrich2007,Assmann2009}. Since the CNLs under test exhibit coherence times of the order of one to ten picoseconds, we expect to measure a significant hint of photon bunching only with detectors having a response time on the same order of magnitude \cite{Vyshnevyy2018}.  Fortunately, superconducting nanowire single-photon detectors (SNSPDs) offer a free-running technique for recording photon arrival times at telecommunication wavelengths with a high detection efficiency and a timing-jitter of only a few tens of picoseconds \cite{Natarajan2012}.

For deconvolution of the measured raw $\smash{g^{(2)}(\tau)}$ histograms, we follow the procedure introduced in Ref.~\cite{Kreinberg2017}. When comparing the coherence time of the emitted light ($\approx$\SI{1}{\pico\second}) to the $\smash{g^{(2)}(\tau)}$-setup response time ($\approx$\SI{80}{\pico\second}) consisting of two SNSPDs, we have to expect a sharply spiking $\smash{g^{(2)}(\tau)-1}$ function that is convolved with a broad instrument impulse response function, loosing almost all information on the original qualitative shape of $\smash{g^{(2)}(\tau)}$.
Therefore, we do not try to fit a model function to the measured data, but instead, we take advantage of the fact that its area $\smash{\tau_\mathrm{area}}=\smash{\int_{-\infty}^{\infty}(g^{(2)}(\tau)-1)\,\mathrm{d}\tau}$  is preserved in the measurement. We extract this area by integrating $\smash{g^{(2)}(\tau)-1}$ over a \SI{320}{\pico\second} window centered around $\smash{\tau}$~=~0.
Assuming that the Siegert relation can be extended to partly thermal light as $\smash{ g^{(2)}(\tau)=1+a| g^{(1)}(\tau)|^{2} }$ with the thermal light fraction $a$, we obtain an estimate for the deconvoluted  $\smash{g^{(2)}(\tau)}$ by dividing the area by the coherence time $\smash{\tau_\mathrm{coh}}=\smash{\int_{-\infty}^{\infty}| g^{(1)}(\tau)|^2\mathrm{d}\tau}$ and adding $1$: $g^{(2)}_\mathrm{deconv}(0)=1+a=1+\frac{\tau_\mathrm{area}}{\tau_\mathrm{coh}}$. We note that the errors of the deconvoluted values of $\smash{g^{(2)}_\mathrm{deconv}(0)}$ increase with decreasing excitation intensity due to the growing spectral width causing shorter temporal coherence of the lasing modes. Additionally, at higher excitation intensities values as low as $\smash{g^{(2)}_\mathrm{deconv}(0)}=1.05$ can be reached as a indication of low excess noise.

\subsection*{Impact of spectral filtering on the measured autocorrelation function}

In order to address the impact of spectral filtering on the autocorrelation measurement of $\smash{g^{(2)}(\tau = 0)}$ of a nanolaser, the nanolaser emission is spectrally filtered to a FWHM of \SI{22}{\pico\meter} (see SI) before the HBT measurement. Raw $\smash{g^{(2)}(0)}$-values as high as $1.85\pm 0.04$ are measured at the lowest excitation intensity. The slight deviation of this result from the ideal value of $\smash{g^{(2)}(0)= 2}$ can be attributed to the temporal resolution of our $\smash{g^{(2)}(\tau)}$-setup. In the studied case, the measured raw $\smash{g^{(2)}(0)}$ values reduce only to approximately 1.67. The measured values of $\smash{g^{(2)}(0)}$ can only decrease with increasing excitation intensity, if the narrowing of the emission line leads to a linewidth comparable to or smaller than the spectral selection.  This ``pseudo-thermal characteristics'' may be misinterpreted as an indication that the nanolaser does not completely undergo the lasing transition. Actually $\smash{g^{(2)}(0)} > 1$ after tight spectral filtering originates from first-order coherence times shorter or similar to the spectral filter time constant resulting in random constructive/destructive interference at the filter output. Such effects have been observed to arrive from the inevitable phase noise of the spontaneous emission~\cite{Neelen1993}, explaining the bunching below \SI{50}{\kilo\watt\per\square\centi\meter} pump intensity. At higher pump intensities the lack of first-order coherence may also be attributed to inhomogeneous broadening, which would agree with the observed change in the spectral lineshape in the region where $\mu < 1$. 




\noindent \textbf{Acknowledgments} The research leading to these results has received funding from from the European Research Council (ERC) under the European Union's Seventh Framework ERC Grant Agreement No. 615613.
K.L. acknowledges support from the Austrian Science Fund (FWF): J-4125-N27.
C.G. and F.L. acknowledge Funding from the Deutsche Forschungsgemeinschaft (DFG) via the graduate school “Quantum-Mechanical Material Modeling” GRK\,2247/1-1.
We thank Marcel Hohn, Nicole Srocka and Tobias Heindel for laboratory support.
\\
\noindent \textbf{Author Contributions}
S.R. and M.K. initiated the research and proposed the experiments. P.H., S.K., K.L., and S.R. are responsible for the  the (quantum-) optical experiments. F.L. and C.G. performed the quantum-optical theoretical analysis.  W.H. and M.K. designed and fabricated the metal clad laser structures. Optical mode simulation was performed by W.H. and M.K. All authors participated in writing the manuscript.
\\
\noindent \textbf{Competing Interests} The authors declare that they have no competing financial interests.
\\
\noindent \textbf{Correspondence} Correspondence and requests for materials should be addressed to Christopher Gies (email: gies@itp.uni-bremen.de), Mercedeh Khajavikhan (email: mercedeh@creol.ucf.edu) and Stephan Reitzenstein (email: stephan.reitzenstein@physik.tu-berlin.de).
\\
\bigskip \noindent See \href{link}{Supplement 1} for supporting content.


\end{document}